\documentclass{article}

\usepackage{microtype}
\usepackage{graphicx}
\usepackage{subfigure}
\usepackage{booktabs} % for professional tables
\usepackage{cprotect}
\usepackage{amsmath, amssymb}
\usepackage{listings}

\usepackage{url}

\usepackage{hyperref}

\usepackage[accepted]{icml2021}

\icmltitlerunning{Optimizing Large Language Models to Expedite the Development of Smart Contracts}

\begin{document}

\setlength{\abovedisplayskip}{0pt}
\setlength{\belowdisplayskip}{2pt}
\setlength{\abovedisplayshortskip}{0pt}
\setlength{\belowdisplayshortskip}{2pt}

\twocolumn[
\icmltitle{Optimizing Large Language Models to Expedite the Development of Smart Contracts}



\icmlsetsymbol{equal}{*}

\begin{icmlauthorlist}
\icmlauthor{Nii Osae Osae Dade}{mai}
\icmlauthor{Margaret Lartey-Quaye}{mai}
\icmlauthor{Emmanuel Teye-Kofi Odonkor}{mai}
\icmlauthor{Paul Ammah}{mai}

\end{icmlauthorlist}

\begin{center}
    Mazzuma Research
\end{center}

\icmlaffiliation{mai}{Mazzuma Research}

\icmlcorrespondingauthor{Nii Osae}{stark@teamcyst.com}

\icmlkeywords{Machine Learning, Smart Contract Code Generation}

\vskip 0.15in
]



\printAffiliationsAndNotice{}  % leave blank if no need to mention equal contribution

\begin{abstract}
 Programming has always been at the heart of technological innovation in the 21st century. With the advent of blockchain technologies and the proliferation of web3 paradigms of decentralised applications, smart contracts have been very instrumental in enabling developers to build applications that reside on decentralised blockchains. Despite the huge interest and potential of smart contracts, there is still a significant knowledge and skill gap that developers need to cross in order to build web3 applications. In light of this, we introduce MazzumaGPT, a large language model that has been optimised to generate smart contract code and aid developers to scaffold development and improve productivity. As part of this research, we outline the optimisation and fine-tuning parameters, evaluate the model’s performance on functional correctness and address the limitations and broader impacts of our research.
\end{abstract}

\vspace{5mm}
\section{Introduction}
\label{sec:intro}

Artificial intelligence has become one of the pioneering tools of innovation within the the world of technology. With the success of deep learning and high performance architectures such as transformers, the machine learning community has become rife with NLP-based innovations. These are primarily as a result of the proliferation of large-language models which produce remarkable results on tasks such as text completion, text summary, composition, sentiment analysis etc. 
With the release of GPT3 \citep{brown2020gpt3} a 175B model and its successor, GPT4 \citep{openai2023gpt4} a multimodal model that performs tasks at the state-of-the-art benchmark level, the machine learning community has seen a flurry of open source models that have been objectively designed towards natural language processing, understanding and reasoning. Within the realm of evaluation, benchmarks such as GLUE \citep{wang2019glue} and SuperGLUE \citep{wang2020superglue} are used to measure the accuracy and performance of NLP-based models.
Besides language tasks, the research community has been tackling the problem of AI-assisted code generation. CodeParrot by Huggingface was released and open sourced based on GPT2 \citep{radford2019gpt2}. Salesforce also released CodeT5 \citep{wang2021codet5} to accelerate the pace of code generation research. Alphacode \citep{Li_2022} by Deepmind is another model that addresses the AI-assisted code generation task. \citet{chen2021codex} also released Codex, a successor to GPT3 \citep{brown2020gpt3} which was trained on open source code retrieved from GitHub and subsequently deployed as GitHub Copilot to assist with code completion and refactoring within code editors and IDEs. The Codex model is able to generate code based on prompts, debug existing code, write simple functions and applets, and explain functionality of code. In light of all these advancements, these models seem to lack proficiency in generating smart contract code which is used in the development of decentralised applications on blockchain platforms. In view of this gap, the objective of this research is to optimise large language models to aid in the generation of smart contract code which will help developers to scaffold and create robust dApps with ease and efficiency.
In this paper, we shall delineate how we fine-tuned a large-language model to produce a custom model (MazzumaGPT) that generates smart contract code, the implementation setup and the model’s performance in comparison with another state-of-the-art model.
Also, review was done on the code generated by the model, assessment of the model’s limitation, concluding with highlights of related work and broader impacts of this research.

\section{Data Collection}
\begin{table*}[t]
\caption{The table below comprises of the various curated use-cases and their respective percentages  that contribute to the training dataset.}
\label{tab:mazzumagpt-dataset}
\vskip 0.15in
\begin{center}
\begin{small}
\begin{sc}
\begin{tabular}{lcc}
\toprule
Use Case Category & Percentage of Dataset \\
\midrule
Frequently Used Smart Contracts & 11.11\%  \\
Programming Data Structures and Alogrithms & 24.73\%  \\
Ethereum Virtual Machine (EVM) based functions and implementations & 27.61\%  \\
DeFi Applications & 24.73\%  \\
Plutus Implementations & 11.82\%  \\
\bottomrule
\end{tabular}
\end{sc}
\end{small}
\end{center}
\vskip -0.1in
\end{table*}

Training data was collected from open source projects written in Solidity and Plutus; two programming languages which are heavily used in the Ethereum and Cardano ecosystems respectively. The code samples contain varying distinct data structures and algorithms. These further broaden the scope of the use cases as seen in Table \ref{tab:mazzumagpt-dataset}\ for which the model can learn, modify and implement to solve a wider variety of problems which may reside outside the training data domain space.\footnote{Full breakdown of the categories can be found in Appendix \ref{sec:use-cases}} Each code sample went through screening and validation to ensure high levels of objectivity and functionality before being added to the training dataset.

\section{Implementation Setup and Training Methods}

Fine-tuning was done using OpenAI’s API since it provided a highly abstracted mechanism to perform the process. Data cleaning and preprocessing was done using the OpenAI data preparation tool as indicated in their documentation. The training dataset comprised of prompt and completion pairs which were sanitised and stored in jsonl format. For performance evaluation, a validation dataset was extracted from the training dataset in order to conduct model performance analysis after training has been completed. The Davinci 175B-parameter model was chosen as the main base model for the fine-tuning process. This base model was chosen due to its ability to understand context and generate outputs which are very accurate to the training data. Regarding training parameters, we used the default batch size along with a variation of different hyperparameter values for each training run to ascertain the right combination of parameters which will achieve the best results for the model.

\section{Results and Evaluation}
The outcome after running several training procedures indicated a linear rise in model performance in relation to model size. This is in conformance with the scaling laws experiments conducted by \citet{kaplan2020scaling}. The use of experimental heuristics was very instrumental in reaching convergence to obtain a model that generates functionally correct code. In the instance of training the Davinci model on 6 epochs with a learning multiplier rate of 0.2, the training token accuracy and training sequence accuracy after fine-tuning were both 1.0 indicating a high accuracy rate of the model's score in generating smart contract code which satisfies the requirements of the prompt.

Due to the niche nature of smart contracts, there wasn't any readily available Solidity or Plutus code benchmark for large-language models. The HumanEval benchmark by \citet{chen2021codex} was tailored for Python code and hence was not appropriate to benchmark a model that generates smart contract code. In this respect, we modified a sample of the problems enumerated in HumanEval to fit the solidity programming language. Majority of the problems involved commonly used data structure and algorithm challenges which should be solvable by any sufficiently advanced code-generating large-language model.

\begin{table}[t]
\caption{The table below shows the performance of MazzumaGPT in comparison to ChatGPT when hand-graded on 10 samples from each model.}
\label{tab:mazzumagpt-eval}
\vskip 0.15in
\begin{center}
\begin{small}
\begin{sc}
\begin{tabular}{lcc}
\toprule
Model & pass@1 \\
\midrule
MazzumaGPT & 80\%  \\
ChatGPT & 70\%  \\
\bottomrule
\end{tabular}
\end{sc}
\end{small}
\end{center}
\vskip -0.1in
\end{table}

In evaluating our model, we took a qualitative approach as suggested by \citet{gunasekar2023textbooks}
as well as the quantitative pass@$k$ method \citep{chen2021codex} where a problem is considered solved if any of the $k$ code samples passes the unit test. In the qualitative approach, we assess a model's coding skills by comparing the similarity of its output to the correct expected solution. This is akin to how developers are assessed during coding interviews. This approach gives insight into the reasoning steps and how the model follows the correct logic to arrive at the solution instead of just relying on the binary results of whether the solution passed the test or not. There are some instances where a model might follow the correct steps but arrive at a wrong solution due to a minor error. In the same way, a model might get the solution right by coincidentally passing the unit test using an inappropriate approach that does not generalise well.

The test comprised of a randomly sampled coding challenge from the modified HumanEval dataset. 10 samples were provided by each model on one attempt (pass@1) where the samples were graded according to the expected solution. 8 out of the 10 samples from the 175B-parameter MazzumaGPT model passed the test while 7 out of 10 samples from ChatGPT passed the test.The final evaluation was done by hand-grading due to the lack of automated ground-truth evaluation on the solidity-based problems.

\begin{figure}[h!]
\centering
\includegraphics[width=\columnwidth]{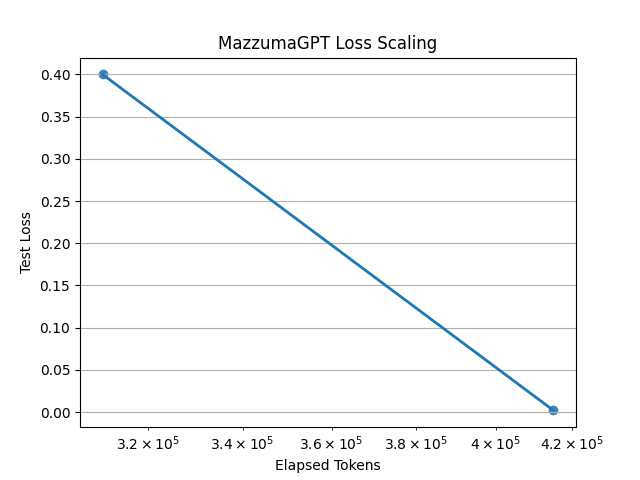}
\vspace*{-2mm}
\caption{MazzumaGPT's cross-entropy loss shows a smooth power of law of scaling performance as the 
number of elapsed tokens are increased.}
\label{fig:loss-scaling}
\vspace*{8mm}
\end{figure}

During the fine-tuning process, we noticed the decrease in cross-entropy loss whenever the model is exposed to more tokens (Figure \ref{fig:loss-scaling}). This was an indicator that performance of the model correlated with the amount of training data. This correlation was also identified by \citet{hoffmann2022training} where model performance was seen to be significantly improved by increasing the training data. Though the model parameter scaling laws experiment by \citet{kaplan2020scaling} still holds ground, the scenario where scaling the elapsed tokens improves performance, colloquially referred to as the Chinchilla hypothesis \citep{hoffmann2022training}, proves to be efficient where model parameter size is kept to an optimal minimum to improve performance during inference.

\begin{figure}[h!]
\centering
\includegraphics[width=\columnwidth]{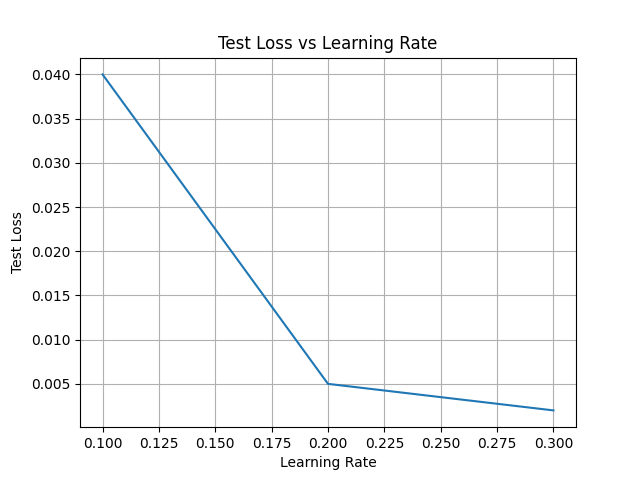}
\vspace*{-2mm}
\caption{In the figure above, it is evident that the test loss reduces as we increase the learning rate. This is an indication that the model's performance improved to the point of convergence.}
\label{fig:loss-vs-learning-rate}
\vspace*{8mm}
\end{figure}

Experimenting with different learning rates provided insight into how the learning rate hyperparameter affects performance for each epoch. As seen in Figure \ref{fig:loss-vs-learning-rate}, the loss dropped as the learning rate was increased to the point of optimal performance. With diverse datasets like what we prepared, we observed that higher learning rates improved performance. However, increasing the learning rate beyond the optimal threshold can also lead to overfitting and consequently degrade performance. One should note that parameters such as dataset or batch size, learning-rate, training epochs and other hyperparameters should be adjusted appropriately to get the best results.

\begin{figure}[h!]
\centering
\includegraphics[width=\columnwidth]{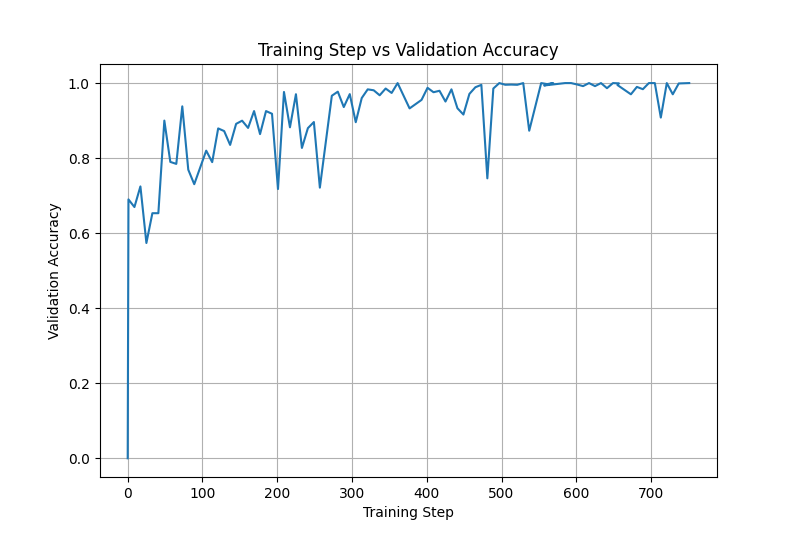}
\vspace*{-2mm}
\caption{Validation accuracy metrics stabilized after 550 training steps, indicating satisfactory model performance on the validation dataset.}
\label{fig:training-step-vs-validation-accuracy}
\vspace*{8mm}
\end{figure}

Also, increasing the training steps resulted in an improvement in validation test accuracy. As seen in Figure \ref{fig:training-step-vs-validation-accuracy}, the validation accuracy score stabilized between $0.97$ - $1$ after 550 training steps. Beyond this point, any further training might result in overfitting and reduce performance.

\begin{figure}[h!]
\centering
\includegraphics[width=\columnwidth]{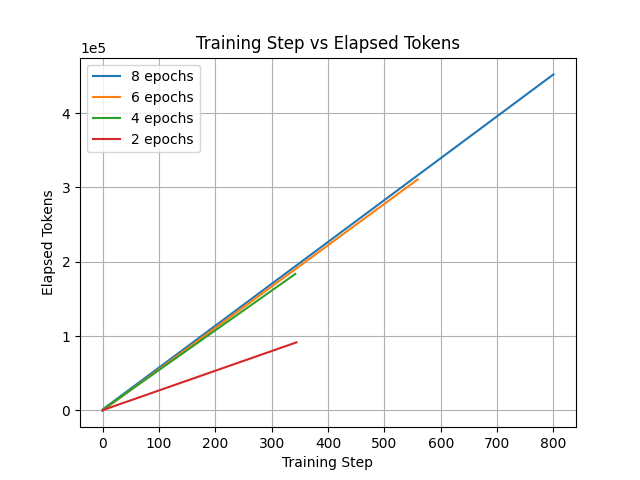}
\vspace*{-2mm}
\caption{The number of elapsed tokens increased linearly with training steps, showing a significant increment in tokens after every 200 steps.}
\label{fig:training-step-vs-elapsed-tokens}
\vspace*{8mm}
\end{figure}

In our experimentation, we also noticed a correlation between the number of training steps (which is directly increased by the number of epochs) and the number of elapsed tokens. Figure \ref{fig:training-step-vs-elapsed-tokens} illustrates this relationship which further buttresses the connection between the elapsed token increment and overall model performance observed in Figure \ref{fig:loss-scaling}. \newline

Due to the varied nature of the dataset that was used, the model can extrapolate to generate more than 1 million unique code samples with diverse functionalities and coding patterns. This ability was contingent on data diversity and inclusion of fundamental data structures and algorithms which are the building blocks of most modern day applications.

On inference, we observed that 0.8 was the optimal temperature to use for code generation. This is coherent with similar observations from \citet{chen2021codex} during the Codex experimentation. Conventionally, temperature parameters for deterministic tasks such as code generation are often closer to zero. For other natural language tasks such as text generation in the category of copy-writing and poetry, the temperature is often closer to one. This parameter change modulates the probability of the model's output to make it either deterministic or creative. In the case of code generation, a temperature of $0.3$ - $0.4$ would have been ideal theoretically. However, 0.8 produced better results due to the activation of in-context learning at higher temperatures which enables the model to creatively adapt and generate a solution that is coherent with the prompt.

\section{Code Analysis and Performance Review}
Here, we will look at how the model interpreted different prompts and provided results that met the requirements of the problems presented.

\begin{figure}[h!]
    \centering
    \begin{lstlisting}[breaklines=true,basicstyle=\ttfamily\scriptsize]
//SPDX-License-Identifier: MIT
pragma solidity^0.8.17 ;
contract HelloWorld{
string public greet = 'Hello World!' ;
} 
\end{lstlisting}
    \caption{Code generated using "Hello world" prompt}
    \label{fig:hello world}
\end{figure}

In Figure \ref{fig:hello world}, the example prompt was not included in the training data; however, the model was able to parse the prompt and write a custom contract that outputs the string within the prompt. This form of in-context learning and inference is often observed in large language models with significantly large parameters. Due to the in-context learning which was inherently part of the pre-trained base model (Davinci 175B), the fine-tuned model (MazzumaGPT) is able to infer solutions and solve problems which it hasn’t seen before.

\begin{figure}[h!]
    \centering
    \begin{lstlisting}[breaklines=true,basicstyle=\ttfamily\scriptsize]
// SPDX-License-Identifier: MIT
pragma solidity = 0.7.6 ;

import '@openzeppelin/contracts/token/ERC721/ERC721.sol';

contract SampleNFT is Ownable, ERC721('NFT_NAME', 'NFT_SYMBOL'){
uint tokenId;
mapping(address=>tokenMetaData[]) public ownershipRecord;
struct tokenMetaData{
uint tokenId;
uint timeStamp;
string tokenURI;
}
function mintToken(address recipient) onlyOwner public {
require(owner()!=recipient, 'Recipient cannot be the owner of the contract');
_safeMint(recipient, tokenId);
ownershipRecord[recipient].push(tokenMetaData(tokenId, block.timestamp, 'ASSET_URL'));
tokenId = tokenId + 1;
}
}
\end{lstlisting}
    \caption{Code generated using "Generate NFTs using ERC721 standard" prompt}
    \label{fig:NFT code}
\end{figure}

The code snippet in Figure \ref{fig:NFT code} creates a baseline scaffold for the user to mint non-fungible tokens using the ERC721 standard. Here, the model generated the code using the library which was specified in the prompt.

\begin{figure}[h!]
    \centering
    \begin{lstlisting}[breaklines=true,basicstyle=\ttfamily\scriptsize]
    
// SPDX-License-Identifier: MIT
contract TestChainlinkOracleParams is  ChainlinkOracle {
address public root = 0xB17314fD8bDe81646a4B2bAC2DD4Bb88Eac360; 
IERC20  public immutable token;
uint public price;
// Call contract inside contract
function getPrice () public view returns (uint){
return  price ;
}
// Call contract outside contract
function getTokenPrice () public view returns (uint) {
return  token.priceOf(this) ;
}
function test () public {
uint bal = getTokenBalance(msg.sender);
uint price = getTokenPrice();
// You can pass address of this to update the price
(uint newPrice) = updatePrice(price);
assert(newPrice == price) ;
assert(bal == token.balanceOf(msg.sender));
}
} 
\end{lstlisting}
    \caption{Prompt:"Write a contract to get a token price using chainlink oracle"}
    \label{fig:chainlink oracle code}
\end{figure}

In Figure \ref{fig:chainlink oracle code}, the model provided a comprehensive solution to the problem by taking a systematic approach which involved creating an individual contract along with a sub-contract to meet the requirements. It also provided in-line comments that explain each function. In addition to this, it also wrote test functions for the user to validate and verify the contracts and functions which have been generated. We also observed that the model adopted the test-driven software engineering design pattern in solving this problem. \newline

\begin{figure}[h!]
    \centering
    \begin{lstlisting}[breaklines=true,basicstyle=\ttfamily\scriptsize]
    
// SPDX-License-Identifier: MIT
pragma solidity ^ 0.8.17 ;
contract MazzumaSwap {
address private constant FACTORY = 0x5C69bEe701ef814a2B6a3EDD4B1652CB9cc5aA6f;
address private constant ROUTER = 0x7a250d5630B4cF539739dF2C5dAcb4c659F2488D;
address private constant WETH = 0xC02aaA39b223FE8D0A0e5C4F27eAD9083C756Cc2;
function swap (address _tokenA, address _tokenB, uint _amountA, uint _amountB) external {
IERC20 (_tokenA).transferFrom(msg.sender,  address(this), _amountA);
IERC20 (_tokenB).transferFrom(msg.sender,  address(this), _amountB);
IERC20 (_tokenA).approve(ROUTER _amountA);
IERC20 (_tokenB).approve(ROUTER _amountA);
(uint amountA, uint amountB,  uint profit) = IUniswapV3Router( ROUTER ).swap( _tokenA,_tokenB,_amountA,_amountB, 1 ,1 ,address(this)block.timestamp);
}
}
---
\end{lstlisting}
    \caption{Prompt:"Create swap contract for mazzuma using uniswap v3 protocol"}
    \label{fig:mazzuma swap code}
\end{figure}

In the Figure \ref{fig:mazzuma swap code} example\footnote{The code samples have been truncated for the purpose of clarity to reduce verbosity. Full code samples can be found in Appendix \ref{sec:generated code}}, the model understood the context and named the contract according to the specifications provided in the prompt. It also implemented the uniswap v3 protocol, which is a very popular token swapping protocol that is used for decentralised finance (DeFi) applications. With this, any developer who is familiar with DeFi can conveniently create DeFi applications without starting from scratch.

\section{Broader Impacts}
\subsection{Limitations}
From the observations that were made, the model does not produce very accurate results on prompts that are too lengthy due to constraints on the context length of the model. Also, even though the code produced functionally met the requirements, some minor errors such as missing commas and semi-colons were seen in the generated code. Hence, it is advised that further inspection, cleaning and testing should be done on generated code before being deployed into production. Code generated by machine learning models should not be taken as a source of ground-truth due to the probability of error occurrence. The purpose of the model is to serve as a sufficiently advanced smart contract co-developer to assist developers in their day-to-day tasks. Furthermore, this model is treated as a productivity tool to enable developers try out different variations of ideas without writing out each iteration manually.

\subsection{Security Concerns}
While curating the training data, code which contained any form of exploits, hacks or vulnerabilities were exempted from the dataset. This was done to prevent any malicious actors from using the model to develop new hacking tools which will pose a threat to the web3 ecosystem. As part of sanity checks that were done, each code sample was audited thoroughly to ensure that the model was fed with only clean code which was free from any form of exploit injection. Besides publicly available smart contract addresses which are used by developers within the space, no private address data or PII (Personally Identifiable Information) was included in the training data.

\subsection{Ethical Concerns and AI Alignment}
Artificial intelligence can be classified as a tool that can have both positive and negative effects depending on the use case and the actors behind the system. Even though thorough measures and standards are observed during data collection, training, evaluation and serving of these models, it is encouraged that extra scrutiny is applied to AI-generated systems that handle sensitive data and applications such as finance, identity and privacy.

\subsection{Bias and Representation}
Due to the fact that the code generation model was fine-tuned on a large-language model, it has the inherent likelihood to exhibit some amount of bias towards its training data. This may lead to the observation of some model outputs that are closely similar to the training data and might prove inefficient if being used for other tasks which are out of the scope of this research. Also, the web3 development community has other languages which are actively being used by developers, and the lack of sufficient training data from these languages may lead to marginalising these communities through unfair representation and inclusion.

\subsection{Environmental Impacts}
Though fine-tuning of large-language models costs a lot of computational power and increases the carbon footprint of the AI value chain, we believe that code generation applications provide a cost-efficient way to develop applications since lesser compute power will be needed to build applications from ground up.
This capability can be scaled to broaden the ability of the model without pre-training and hence saves up compute power for those who will want to improve on our work.

\subsection{Intellectual Property}
The training data that was used in this research was under the MIT license that allows modification, reproduction and redistribution of the code. This conforms with the open source ethos of software production and distribution. Hence, any code generated by the model falls under the MIT license category. We believe this will give developers the leverage to build and use resources to benefit the entire open source community and proliferate the growth of the open source software.

\subsection{Economic Impacts}
Due to the increase in code generation tools, we believe it will drive down the time to produce software applications and hence reduce the cost of production within the software value chain.
Even though the timeline for this change is not certain, there will be a gradual increase in the perception that AI-powered program synthesis might replace developers. Due to the nature of modern software development, we believe these tools will augment and enhance the work of developers rather than replace them. As code generation tools become more powerful and sophisticated, they will improve the productivity of software development and enable them to create innovative and novel solutions which wouldn’t have been practically feasible without the help of artificial intelligence.

Furthermore, we intend to open source the training data of our research and encourage the developer community to contribute to the dataset. Each developer’s contribution will go through an approval process before being added to the master dataset. Upon approval, the developer will be rewarded with community tokens which can be used to generate smart contract code from the model. This mechanism design is in line with the ethos of data decentralisation which is very common in the blockchain/web3 ecosystem. As the community continues to contribute to the dataset, the code generation model will be able to scale significantly to meet the ever growing demands of the ecosystem. We believe this mechanism will lead to sustainable growth and maintenance of the project.

\section{Related Work}
\subsection{Program Synthesis}
Prior to the explosion of the wide usage of deep neural networks \citep{DBLP:journals/cm/LeCunJBDGGHHH89}, program synthesis has been a topic that has accumulated a lot of research within the artificial intelligence space. The deductive synthesis approach \citep{manna1971progsyn} where specifications are converted into constraints which are then passed into a theorem prover that derives a proof that satisfies the constraints that have been specified. 
More recently, the use of deep learning architectures such as recurrent networks were used by \citet{DBLP:journals/corr/Neubig17} to produce code by using attention mechanism in mapping of text to abstract syntax trees. Using a program induction approach, \citet{zaremba2014learning} worked on models that could take on trivial tasks such as memorisation and addition using latent program representation. Other implementations such as the Neural Program Interpreter \citep{reed2016neural,shin2018improving,pierrot2021learning}, the Neural GPU \citep{kaiser2015neural}, the Universal Transformer \citep{dehghani2019universal} and the memory networks \citep{weston2015memory,sukhbaatar2015endtoend} have observed significant progress within the program induction domain.

Alternative approaches were also taken in the realm of pseudocode conversion to code \citep{kulal2019spoc}, generation of program sketches \citep{guo2022learning,murali2018neural} , reinforcement learning domain generation of programmatic policies \citep{trivedi2022learning} and guided program search by \citet{balog2017deepcoder}. 
Automated completion of code has grown to become an important part of modern software development especially with integrated development environments (IDEs) and code editors \citep{Li_2022}. Code completion tools suggest possible continuations for the code that is being typed into the interface and majority of the earliest tools were purely syntax-based \citep{Li_2022}. \citet{hindle2012naturalness} worked on n-gram language models of code and this indicated that sequence of code was more predictable than natural language. \citet{allamanis2015code} incorporated the idea of learning a state vector used to condition child node propagation to implement a text-to-code retrieval system. \citet{DBLP:journals/corr/Neubig17} also applied it in text-conditional code generation.
 According to \citet{chen2021codex}, in program synthesis, a model explicitly generates a program mostly from specifications written in natural language. Among many classical approaches, the most popular is using probabilistic context-free grammar (PCFG) to generate a program’s abstract syntax tree (AST). Majority of the aforementioned implementations resorted to classical search and next-word prediction algorithms to synthesise programs that are in adherence to formal specifications which are accorded to each generation task. In order to explore other approaches, \citet{AlphaDev2023} used deep reinforcement learning to discover faster sorting algorithms.

\subsection{Architecture for Code Generation}
After demonstrating the potential and success of large-scale transformers to natural language processing and modeling \citep{brown2020gpt3}, this propelled research into the use of transformer models for code translation, retrieval and generation \citep{chen2021codex,clement2020pymt5,feng2020codebert}. These models showed outstanding results in text generation. Further work was done by training the Generative Pretrained Transformer (GPT) language model \citep{radford2019gpt2} on public code from GitHub. The model, named Codex \citep{chen2021codex}, produced stunning results in code generation in Python with specification from docstring. The production versions of this model was further trained on other programming languages and released as GitHub Copilot, an assistive programming tool powered by artificial intelligence. PyMT5 \citep{clement2020pymt5} used the T5 objective to train a system which can translate between non-overlapping subsets of signature, docstring and body of code. This work has a methodological resemblance to the research done by \citet{chen2021codex}. Similar work was also done by \citet{austin2021program} who demonstrated that fine-tuning a model on programming task dataset can improve the success rate when given other tasks within similar domains. \citet{nijkamp2023codegen} also demonstrated similar results using the JAXformer training library to release a family of models that generate code from prompts. \citet{trummer2022codexdb} also investigated on SQL based code generation using GPT-3 Codex. In order to augment code generation using feedback loops, \citet{coderl2022} worked on CodeRL which uses deep reinforcement learning to create an actor and critic framework to improve the quality of program synthesis.
Regarding work done on code exclusive models, a team at Meta created an internal programming assistant called CodeCompose \citep{murali2023codecompose} using the InCoder LLM \citep{fried2023incoder}. Also, \citet{li2023starcoder} used the Nvidia Megatron-LM \citep{shoeybi2020megatronlm} framework to create Starcoder, a 15.5B parameter model trained on 1 trillion tokens of code. To further improve this, \citet{luo2023wizardcoder} used the evol-instruct method from \citep{xu2023wizardlm} to fine-tune Starcoder and create WizardCoder, which outperformed most open source models by a substantial margin.

\subsection{Evaluation Metrics and Benchmarks}
There have been several benchmarks when it comes to evaluating the performance of large language models. The introduction of BLEU (Bilingual Evaluation Understudy) \citep{wang2019glue} and SuperGLUE \citep{wang2020superglue} in the domain of natural language processing and understanding enabled researchers to compare the performance of large language models. Within the realm of code generation, \citet{ren2020codebleu} indicated that BLEU encounters problems when capturing semantic features that are specific to code and hence calls for modifications to the score. In alignment to develop a benchmark that takes functional correctness into consideration, work by \citet{lachaux2020unsuptrans} and \citet{kulal2019spoc} measures performance based on a metric where a sample is considered correct if it passes a set of unit tests. This evaluation of functional correctness by \citet{kulal2019spoc} using the pass@$k$ metric was also used by \citet{chen2021codex} in their HumanEval benchmark system which measures performance of synthesising programs from docstrings. \citet{Li_2022} used the similar pass@$k$ metric in evaluating performance on competition-level code generation. Also, \citet{nijkamp2023codegen} introduced a Multi-Turn Programming Benchmark (MTPB) to investigate a model’s capacity on synthesising programs in a multi-step paradigm.

\section{Further Discussions and Future Work}
In training MazzumaGPT, we noticed that the quality of data was very crucial to attain model accuracy and functional correctness of generated code. Unlike natural language text generation whereby random text and conversations can be used to improve a model's ability to hold conversations, code generation poses a different kind of challenge. Here, it is essential for the model to understand and parse the prompt by breaking down the problem into smaller functions and then combining these functions to solve the problem presented in the prompt. This manner of divide and conquer approach demands that the model follows a particular set of algorithms or instructions in order to adequately solve a problem. This is where instruction fine-tuning played a significant role in improving model performance using a smaller but quality dataset. This methodology was also attested by \citet{zhou2023lima} where a 65B parameter model which was fine-tuned on only 1,000 curated quality dataset outperformed a lot of open source models on several benchmarks. Using instruction fine-tuning also helps to align the model to the user's intent and create appropriate guardrails against abuse and misuse.
Also, we noticed that training a model on its own synthetically generated data degrades model performance. This phenomenon termed as model collapse \cite{shumailov2023curse} increases the statistical errors of the model, making it forget its original data distribution and further reduce its performance.

Within the scope of our work, we demonstrated that with high quality and well curated data, a large language model can be fine-tuned to generate code with impressive levels of functional correctness. Future work on this will experiment this method on other foundation models to achieve higher degrees of code accuracy and model enhancements.

\section{Conclusion}
In this research, we sought to optimise a large language model to generate smart contract code by fine-tuning the model on Solidity and Plutus code. We investigated the results of using different hyperparameters during the fine-tuning process to reach optimal results. Furthermore, we used the pass@$k$ benchmark to compare our model’s performance in smart contract code generation. We also found that altering the temperature can significantly affect the performance of the model and consequentially the functional correctness of the code generated. In addition to this, we explored the limitations of the model, addressed broader impacts of this research such as security, economic impacts and ethical alignment. Finally, we looked at relevant work within the field and outlined our vision to decentralise our research to scale and improve our model for the community.

\section*{Acknowledgements}
\label{sec:ack}
We will like to thank Kofi Genfi, Brenton Naicker, Gideon Greaves, Yoseph Ayele, Vitalik Buterin, Michiel Bellen, Wei Xiao, Ali Soleman and Derrick Selempo for their thoughtful discussions and feedback.

\small
\bibliography{main}
\bibliographystyle{icml2021}
\normalsize
\appendix
\section{Trained Use Cases}
\label{sec:use-cases}
\textbf{Frequently Used Contracts}
\begin{itemize}
        \item ERC20 token
        \item ERC721
        \item Token Swap
        \item ERC1155
        \item Getting quote from Uniswap
        \item Defi token trading
        \item Uniswap liquidity provision
        \item Swap oracle
        \item Flash loan
        \item Handle ether and wei currency breakdown
\end{itemize}

\textbf{Programming data structures and algorithms}
\begin{itemize}
        \item Conditional statements 
        \item Array data structures
        \item Enumerables 
        \item Structs
        \item Variable storage
        \item Generic functions
        \item View getter
        \item Error handling
        \item Function modifier
        \item Event logging
        \item String output
        \item Manage variable store
        \item Attach params to variables
        \item Data type declarations
        \item Constants declarations
        \item Immutable data declarations
        \item Read and write state variables 
        \item Object-oriented style programming
        \item Inheritance 
        \item Public visibility
        \item Private visibility
        \item Internal functions
        \item External functions
\end{itemize}

\textbf{Ethereum Virtual Machine (EVM) based functions and implementations}
\begin{itemize}
        \item Internal state variables
        \item External State variables
        \item Interface implementation
        \item Payable contract
        \item Fallback contract
        \item Send ether
        \item Callable contracts
        \item Delegatecall contract
        \item Function selector
        \item Contract factory
        \item Try/catch error handling
        \item Simple math library
        \item ABI encode
        \item ABI decode
        \item Keccak-256 hash implementation 
        \item Signature verification
        \item Unchecked maths contracts
        \item Ether wallet contract
        \item Multi-sig wallet contract
        \item Slot assembly contract
        \item Uni-directional payment channel
        \item Create2 implementation
        \item Proxy deploy implementation
        \item Merkle tree implementation
        \item Iterable mapping
        \item Bi-directional payment channel
        
\end{itemize}

\textbf{Defi Applications}
\begin{itemize}
        \item Non-fungible token (NFT) English Auction
        \item Multicall smart contracts
        \item Multi delegate call smart contracts
        \item Time lock contract
        \item NFT Ducth auction
        \item Crowdfunding campaign
        \item Uniswap v2 token swap
        \item Uniswap v2 liquidity addition
        \item Uniswap v2 liquidity removal
        \item Uniswap v2 one-sided supply
        \item Uniswap v2 flash swap
        \item Uniswap v3 token swap
        \item Uniswap v3 position minting
        \item Uniswap v3 fee collection from position 
        \item Uniswap v3 liquidity management 
        \item Uniswap v3 flash loan
        \item Uniswap v3 flash swap arbitrage
        \item Price fetch from Chainlink oracle
        \item Staking smart contract
        \item Discrete staking smart contract
        \item Defi vault implementation
        \item Constant sum automated market maker
        \item Constant product automated market maker

\end{itemize}

\textbf{Plutus Implementations}
\begin{itemize}
        \item Plutus NFT minting 
        \item Message logging
        \item Address monitoring
        \item Plutus native token minting policy 
        \item Burning tokens in plutus
        \item On-chain minting policy
        \item Off-chain minting policy
        \item Generic contract in plutus
        \item Trace tokens in plutus
        \item Check math operations in plutus
        \item Minting policy signing in plutus
\end{itemize}

\section{Prompts With Generated Code Samples}
\label{sec:generated code}
Prompt: Implement a fallback contract in solidity

    \begin{lstlisting}[breaklines=true,basicstyle=\ttfamily\scriptsize]
    
// SPDX-License-Identifier: MIT
pragma solidity ^0.8.17;

contract Fallback {
    event Log(string func, uint gas);

    // Fallback function must be declared as external.
    fallback() external payable {
        // send / transfer (forwards 2300 gas to this fallback function)
        // call (forwards all of the gas)
        emit Log('fallback', gasleft());
    }

    // Receive is a variant of fallback that is triggered when msg.data is empty
    receive() external payable {
        emit Log('receive', gasleft());
    }

    // Helper function to check the balance of this contract
    function getBalance() public view returns (uint) {
        return address(this).balance;
    }
}

\end{lstlisting}

Prompt: write a smart contract for implementing an NFT english auction in solidity

    \begin{lstlisting}[breaklines=true,basicstyle=\ttfamily\scriptsize]
// SPDX-License-Identifier: MIT
pragma solidity ^0.8.17;

interface IERC721 {
    function safeTransferFrom(address from, address to, uint tokenId) external;

    function transferFrom(address, address, uint) external;
}

contract EnglishAuction {
    event Start();
    event Bid(address indexed sender, uint amount);
    event Withdraw(address indexed bidder, uint amount);
    event End(address winner, uint amount);

    IERC721 public nft;
    uint public nftId;

    address payable public seller;
    uint public endAt;
    bool public started;
    bool public ended;

    address public highestBidder;
    uint public highestBid;
    mapping(address => uint) public bids;

    constructor(address _nft, uint _nftId, uint _startingBid) {
        nft = IERC721(_nft);
        nftId = _nftId;

        seller = payable(msg.sender);
        highestBid = _startingBid;
    }

    function start() external {
        require(!started, 'started');
        require(msg.sender == seller, 'not seller');

        nft.transferFrom(msg.sender, address(this), nftId);
        started = true;
        endAt = block.timestamp + 7 days;

        emit Start();
    }

    function bid() external payable {
        require(started, 'not started');
        require(block.timestamp < endAt, 'ended');
        require(msg.value > highestBid, 'value < highest');

        if (highestBidder != address(0)) {
            bids[highestBidder] += highestBid;
        }

        highestBidder = msg.sender;
        highestBid = msg.value;

        emit Bid(msg.sender, msg.value);
    }

    function withdraw() external {
        uint bal = bids[msg.sender];
        bids[msg.sender] = 0;
        payable(msg.sender).transfer(bal);

        emit Withdraw(msg.sender, bal);
    }

    function end() external {
        require(started, 'not started');
        require(block.timestamp >= endAt, 'not ended');
        require(!ended, 'ended');

        ended = true;
        if (highestBidder != address(0)) {
            nft.safeTransferFrom(address(this), highestBidder, nftId);
            seller.transfer(highestBid);
        } else {
            nft.safeTransferFrom(address(this), seller, nftId);
        }

        emit End(highestBidder, highestBid);
    }
}
\end{lstlisting}

Prompt: write a contract for signing a minting policy in plutus

\begin{lstlisting}[breaklines=true,basicstyle=\ttfamily\scriptsize]
    
module SignedContract where

import           Control.Monad          hiding (fmap)
import           Data.Aeson             (ToJSON, FromJSON)
import           Data.Text              (Text)
import           Data.Void              (Void)
import           GHC.Generics           (Generic)
import           Plutus.Contract        as Contract
import           Plutus.Trace.Emulator  as Emulator
import qualified PlutusTx
import           PlutusTx.Prelude       hiding (Semigroup(..), unless)
import           Ledger                 hiding (mint, singleton)
import           Ledger.Constraints     as Constraints
import qualified Ledger.Typed.Scripts   as Scripts
import           Ledger.Value           as Value
import           Playground.Contract    (printJson, printSchemas, ensureKnownCurrencies, stage, ToSchema)
import           Playground.TH          (mkKnownCurrencies, mkSchemaDefinitions)
import           Playground.Types       (KnownCurrency (..))
import           Prelude                (IO, Show (..), String)
import           Text.Printf            (printf)
import           Wallet.Emulator.Wallet

{-# INLINABLE mkPolicy #-}
mkPolicy :: PaymentPubKeyHash -> () -> ScriptContext -> Bool
mkPolicy pkh () ctx = txSignedBy (scriptContextTxInfo ctx) $ unPaymentPubKeyHash pkh

policy :: PaymentPubKeyHash -> Scripts.MintingPolicy
policy pkh = mkMintingPolicyScript $
    $$(PlutusTx.compile [|| Scripts.wrapMintingPolicy . mkPolicy ||])
    `PlutusTx.applyCode`
    PlutusTx.liftCode pkh

curSymbol :: PaymentPubKeyHash -> CurrencySymbol
curSymbol = scriptCurrencySymbol . policy

data MintParams = MintParams
    { mpTokenName :: !TokenName
    , mpAmount    :: !Integer
    } deriving (Generic, ToJSON, FromJSON, ToSchema)

type FreeSchema = Endpoint 'mint' MintParams

mint :: MintParams -> Contract w FreeSchema Text ()
mint mp = do
    pkh <- Contract.ownPaymentPubKeyHash
    let val     = Value.singleton (curSymbol pkh) (mpTokenName mp) (mpAmount mp)
        lookups = Constraints.mintingPolicy $ policy pkh
        tx      = Constraints.mustMintValue val
    ledgerTx <- submitTxConstraintsWith @Void lookups tx
    void $ awaitTxConfirmed $ getCardanoTxId ledgerTx
    Contract.logInfo @String $ printf 'forged %s' (show val)

endpoints :: Contract () FreeSchema Text ()
endpoints = mint' >> endpoints
  where
    mint' = awaitPromise $ endpoint @'mint' mint

mkSchemaDefinitions ''FreeSchema

mkKnownCurrencies []

test :: IO ()
test = runEmulatorTraceIO $ do
    let tn = 'ABC'
    h1 <- activateContractWallet (knownWallet 1) endpoints
    h2 <- activateContractWallet (knownWallet 2) endpoints
    callEndpoint @'mint' h1 $ MintParams
        { mpTokenName = tn
        , mpAmount    = 555
        }
    callEndpoint @'mint' h2 $ MintParams
        { mpTokenName = tn
        , mpAmount    = 444
        }
    void $ Emulator.waitNSlots 1
    callEndpoint @'mint' h1 $ MintParams
        { mpTokenName = tn
        , mpAmount    = -222
        }
    void $ Emulator.waitNSlots 1

\end{lstlisting}

Prompt: write an off-chain token minting policy in plutus

    \begin{lstlisting}[breaklines=true,basicstyle=\ttfamily\scriptsize]
module Token.OffChain
    ( TokenParams (..)
    , adjustAndSubmit, adjustAndSubmitWith
    , mintToken
    ) where

import           Control.Monad               hiding (fmap)
import           Data.Aeson                  (FromJSON, ToJSON)
import qualified Data.Map                    as Map
import           Data.Maybe                  (fromJust)
import           Data.OpenApi.Schema         (ToSchema)
import           Data.Text                   (Text, pack)
import           Data.Void                   (Void)
import           GHC.Generics                (Generic)
import           Plutus.Contract             as Contract
import           Plutus.Contract.Wallet      (getUnspentOutput)
import qualified PlutusTx
import           PlutusTx.Prelude            hiding (Semigroup(..), unless)
import           Ledger                      hiding (mint, singleton)
import           Ledger.Constraints          as Constraints
import qualified Ledger.Typed.Scripts        as Scripts
import           Ledger.Value                as Value
import           Prelude                     (Semigroup (..), Show (..), String)
import qualified Prelude
import           Text.Printf                 (printf)

import           Token.OnChain
import           Utils                (getCredentials)

data TokenParams = TokenParams
    { tpToken   :: !TokenName
    , tpAmount  :: !Integer
    , tpAddress :: !Address
    } deriving (Prelude.Eq, Prelude.Ord, Generic, FromJSON, ToJSON, ToSchema, Show)

adjustAndSubmitWith :: ( PlutusTx.FromData (Scripts.DatumType a)
    , PlutusTx.ToData (Scripts.RedeemerType a)
    , PlutusTx.ToData (Scripts.DatumType a)
    , AsContractError e
            )
        => ScriptLookups a
        -> TxConstraints (Scripts.RedeemerType a) (Scripts.DatumType a)
        -> Contract w s e CardanoTx
adjustAndSubmitWith lookups constraints = do
    unbalanced <- adjustUnbalancedTx <$> mkTxConstraints lookups constraints
    Contract.logDebug @String $ printf 'unbalanced: %s' $ show unbalanced
    unsigned <- balanceTx unbalanced
    Contract.logDebug @String $ printf 'balanced: %s' $ show unsigned
    signed <- submitBalancedTx unsigned
    Contract.logDebug @String $ printf 'signed: %s' $ show signed
    return signed

adjustAndSubmit :: ( PlutusTx.FromData (Scripts.DatumType a)
    , PlutusTx.ToData (Scripts.RedeemerType a)
    , PlutusTx.ToData (Scripts.DatumType a)
    , AsContractError e
    )
    => Scripts.TypedValidator a
    -> TxConstraints (Scripts.RedeemerType a) (Scripts.DatumType a)
    -> Contract w s e CardanoTx
adjustAndSubmit inst = adjustAndSubmitWith $ Constraints.typedValidatorLookups inst

mintToken :: TokenParams -> Contract w s Text CurrencySymbol
mintToken tp = do
    Contract.logDebug @String $ printf 'started minting: %s' $ show tp
    let addr = tpAddress tp
    case getCredentials addr of
        Nothing      -> Contract.throwError $ pack $ printf 'expected pubkey address, but got %s' $ show addr
        Just (x, my) -> do
            oref <- getUnspentOutput
            o    <- fromJust <$> Contract.txOutFromRef oref
            Contract.logDebug @String $ printf 'picked UTxO at %s with value %s' (show oref) (show $ _ciTxOutValue o)

    let tn          = tpToken tp
        amt         = tpAmount tp
        cs          = tokenCurSymbol oref tn amt
        val         = Value.singleton cs tn amt
        c           = case my of
        Nothing -> Constraints.mustPayToPubKey x val
        Just y  -> Constraints.mustPayToPubKeyAddress x y val
            lookups     = Constraints.mintingPolicy (tokenPolicy oref tn amt) <>
                              Constraints.unspentOutputs (Map.singleton oref o)
        constraints = Constraints.mustMintValue val          <>
        Constraints.mustSpendPubKeyOutput oref <>
        c

    void $ adjustAndSubmitWith @Void lookups constraints
    Contract.logInfo @String $ printf 'minted %s' (show val)
    return cs

\end{lstlisting}

\end{document}